\documentclass[
twocolumn,
]{ceurart}

\usepackage{listings}
\usepackage{amsmath}
\begin{document}

\copyrightyear{2022}
\copyrightclause{Copyright for this paper by its authors.
  Use permitted under Creative Commons License Attribution 4.0
  International (CC BY 4.0).}

\conference{ACM SIGKDD Conference on Knowledge Discovery and
Data Mining, AdKDD Workshop 2024, August 2024, Barcelona, Spain}

\title{SERP Interference Network and Its Applications in Search Advertising}

\author[1]{Purak Jain}[%
orcid=0009-0009-9758-9336,
email=purakjn@amazon.com,
url=https://www.amazon.science/author/purak-jain
]
\cormark[1]
\address[1]{Amazon, Seattle, USA}

\author[1]{Sandeep Appala}[%
email=appalar@amazon.com
]

\cortext[1]{Corresponding author.}

\begin{abstract}
  Search Engine marketing teams in the e-commerce industry manage global search engine traffic to their websites with the aim to optimize long-term profitability by delivering the best possible customer experience on Search Engine Results Pages (SERPs). In order to do so, they need to run continuous and rapid Search Marketing A/B tests to continuously evolve and improve their products. However, unlike typical e-commerce A/B tests that can randomize based on customer identification, their tests face the challenge of anonymized users on search engines. On the other hand, simply randomizing on products violates Stable Unit Treatment Value Assumption for most treatments of interest. In this work, we propose leveraging censored observational data to construct bipartite (Search Query to Product Ad or Text Ad) SERP interference networks. Using a novel weighting function, we create weighted projections to form unipartite graphs which can then be use to create clusters to randomized on. We demonstrate this experimental design's application in evaluating a new bidding algorithm for Paid Search. Additionally, we provide a blueprint of a novel system architecture utilizing SageMaker which enables polyglot programming to implement each component of the experimental framework.
\end{abstract}

\begin{keywords}
  sponsored search \sep
  experiment design \sep
  A/B testing
\end{keywords}

\maketitle

\section{Introduction}

Search Engine marketing teams in the e-commerce industry manage global search engine traffic with the aim of optimizing long-term profitability by delivering the best possible customer experience on the most important web pages on the internet - Search Engine Results Pages (SERPs). Figure 1 shows the prominent parts of SERP. Search Engines continue to evolve their customer experience and features due to social, technological and economic forces, including privacy concerns, and further monetization of their properties (SERPs). In anticipation of opportunities and risks that come with a shifting landscape advertisers continuously innovate with new bidding algorithms, improved paid and free search creatives, landing pages etc. Randomized experiments, or A/B tests, are the standard approach for evaluating causal effects of new features \cite{kohavi2013online}. However, Search Marketing experiments are unlike conventional A/B tests in industry that can randomize on customers as advertisers don't identify their customers when they are on a search engine i.e. the ad publisher. Instead, advertisers may run A/B tests randomized by geographic locations \cite{vaver2011measuring} using search engine's geo-targeting capabilities but due to ad publisher's API limitations they are unable to do so without having to clone entire advertisement campaigns. The cloning of entire accounts is operationally expensive and time consuming restricting the velocity at which they can run such trials.

The next obvious choice for unit of randomization is usually products or search queries. However for any A/B test, splits of the unit of randomization should satisfy the assumptions of the Neyman - Rubin causal framework \cite{Rubin2005}, that is, no interference, unconfoundness, overlap and no hidden treatment variations. For example, if we simply randomly select products into control and treatment groups, it should hold the unconfoundness and the overlap assumption given a large sample size but we still need to check if the "no interference" assumption holds. We observe that for most treatments of interest (e.g. new bidding algorithms for paid search programs or improved title headlines for free search snippets), the SERP page leads to interference between treatment and control units causing the Stable Unit Treatment Value Assumption (SUTVA) to fail, and consequently induces bias in the standard estimators used to evaluate the value generated by the treatment. A standard answer \cite{cook2007introduction, bojinov2023design} to this problem is to replace the “product-split” experiment design with a “time-split” (or “switchback”) design, where the entire market switches repeatedly between treatment and control. In practice, such designs turn out to be equally time consuming as geo-based splits since we need to account for long lengths of adjustment period between switches due to the presence of an intermediary i.e. search engine that applies the treatment and takes its own time which advertisers cannot control.

\begin{figure}[h]
  \centering
  \includegraphics[width=\linewidth]{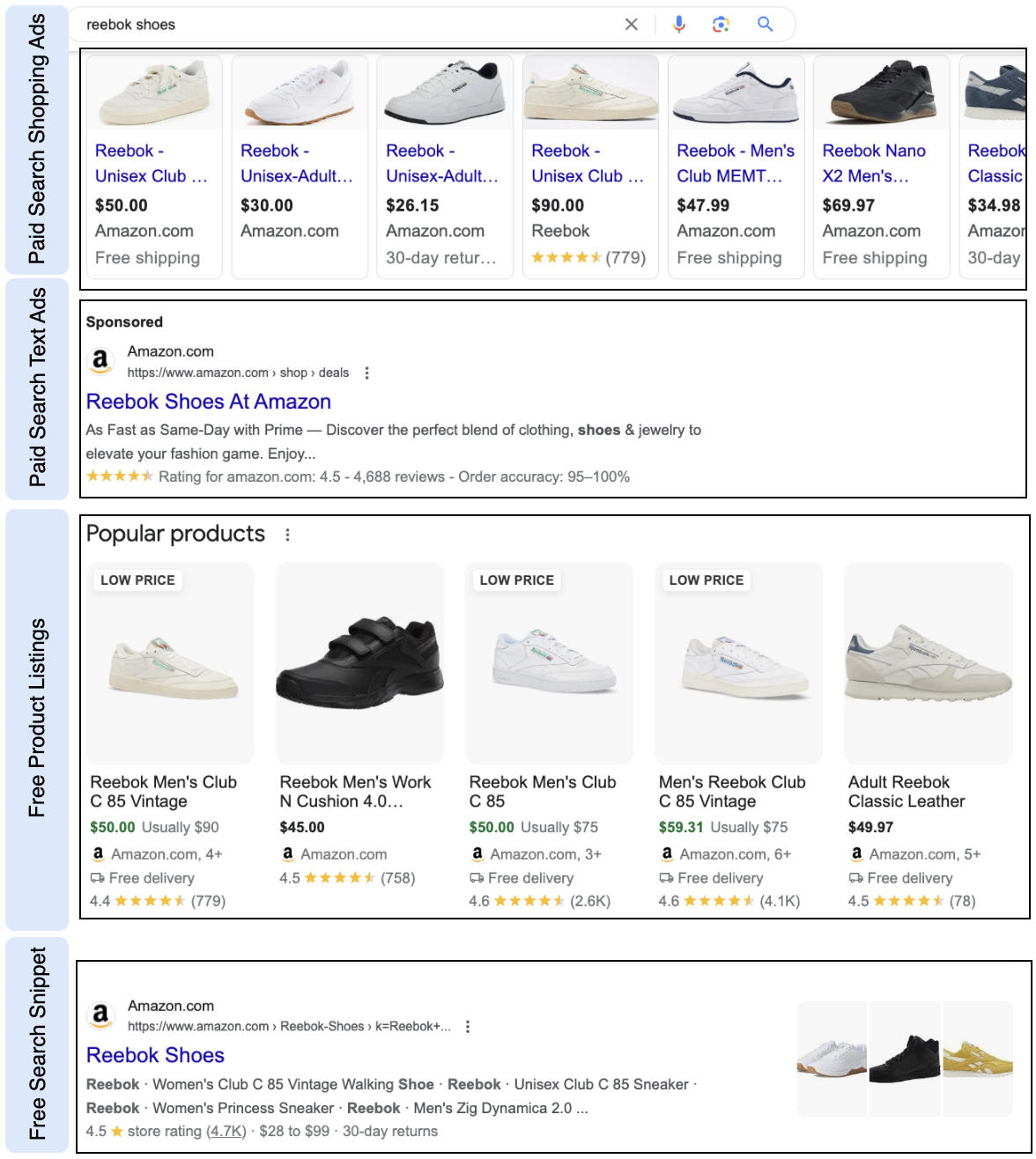}
  \caption{Components of interest on a Search Engine Results Page.}
\end{figure}

Another approach to dealing with spillovers or interference is given by clustered experiments \cite{roberts2005design, aronow2017estimating} or in social network settings, by network bucketing testing \cite{backstrom2011network} where nodes that are relatively clustered together are given the same assignment of treatment or control \cite{ugander2013graph}. Our work is inspired by similar clustered experiments methods that have been applied to estimate and reduce bias in marketplace experiments \cite{holtz2020reducing}. Speciﬁcally, one such example involves experimentation in internet ad auctions, where each auction consists of a keyword along with a set of advertisers who submit competing bids in order for their ads to be displayed when the keyword is queried by a user. There is cross-unit interference because the same advertiser or keyword may appear in multiple auctions. \citet{basse2016randomization} and \citet{ostrovsky2011reserve} make the observation that the auction type used for one keyword does not meaningfully affect how advertisers bid for other keywords. They then consider experiments that group auctions into clusters by their keywords and randomize auction formats across these keyword clusters, rather than across advertisers, as a means to avoid problems with interference. More broadly, in our context of Search Marketing, this idea of cluster-level randomization corresponds to identifying product or search query clusters that are relatively isolated from each other and randomizing the interventions across product-clusters rather than across products. Our primary contribution lies in leveraging observational data to build bipartite (Search Query - Product) and tripartite (Search Query - Paid Search Product - Free Search URL) SERP interference networks. We introduce an innovative weight function to generate weighted projections, transforming these networks into unipartite graphs. These graphs facilitate the clustering of products that co-appear on SERPs through Paid Search Shopping Ads, Text Ads, or Free Product Listings. The resultant clusters can then be randomized during A/B tests to generate insights.

Note that more recently, \citet{johari2022experimental} and \citet{bajari2021multiple} have proposed newer experiment designs where both search query and product units are randomized simultaneously. While having a similar ﬂavor, neither framework applies easily to our problem of interest. To begin with, we cannot control "search-query" assignment as that is determined by the search engine i.e. the ad publisher. \citet{johari2022experimental} use a choice model to capture spillovers, which captures a different kind of market than the one we consider, where interference is mediated by a matching algorithm. \citet{bajari2021multiple} imposes a local interaction assumption, which does not hold in our setting. However, when the graph is a bi-partite graph it holds some similarity which we plan to explore in future work for measuring the magnitude of spillovers. 

The rest of this paper proceeds as follows. In Section 2, we use the two-population search query - product case as a motivation to build SERP interference network to test out new bidding algorithms. In Section 3, we describe in greater detail our experiment design. In Section 4, we further share details on testing a new bidding model using this experimentation design. Section 5 provides an overview of a system architecture blueprint for deploying such experimentation frameworks. Finally, we discuss our ﬁndings and future extensions in Section 6.

\section{Setting and Motivation}

Shopping Ads is one of the ad formats supported on SERPs. To place ads within the shopping ad carousel advertisers need to participate in an auction competing with other advertisers. The format of the auction is considered close to second price Vickrey–Clarke–Groves (VCG) \cite{vickrey1961counterspeculation, clarke1971multipart, groves1973incentives}, although the exact ad publisher implementation is a blackbox for us. As such to maximize long term profitability, it is important to constantly develop, test and launch new bidding algorithms responsible for valuating products worldwide. Let’s say, to test out a new bidding algorithm we simply split on products. The Stable Unit Treatment Value Assumption (SUTVA) presumes that the valuation assigned to one product by the new algorithm does not influence the profitability of other products. However, in the context of shopping advertisements, this assumption may be violated due to potential between-product interference. This interference occurs when both a product with a treatment bid from the new model and another with a control bid from the current model participate in the same auction triggered by a search query, deemed relevant by the ad publisher for both products. See figure 2 for an example. Such scenarios clearly breach SUTVA, challenging the validity of our evaluation method. If one product happens to be assigned to treatment group and the other one to control, then the difference in ﬁnancial performance between the two products will be resulted from the combined effect of treatment and between-product spillover effect, thus making the treatment effect indistinguishable from the product spillover effect.

\begin{figure}[h]
  \centering
  \includegraphics[width=\linewidth]{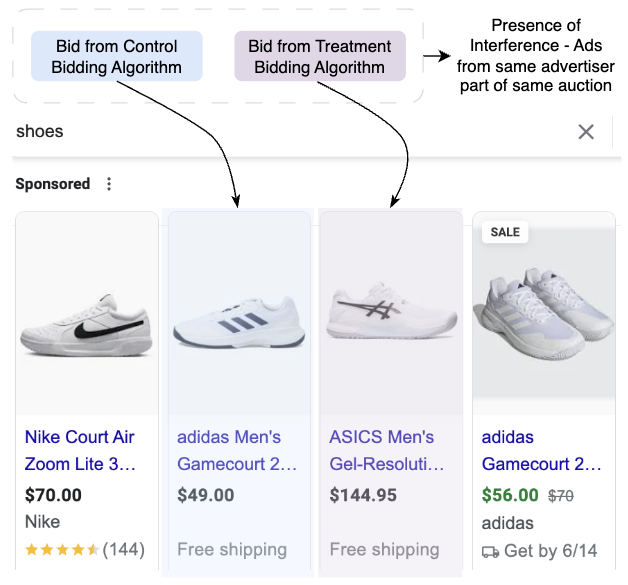}
  \caption{Note the two Shopping Ad advertisements shown in the same carousel. If these two products have bids from different bidders (i.e. control and treatment) then Stable Unit Treatment Value Assumption is violated.}
\end{figure}

\begin{table*}
  \caption{Mock row in a Search Engine's Query Report shared with the advertiser.}
  \label{tab:sample}
  \begin{tabular}{cccccc}
    \toprule
    Search Query & Impressions & Product/Keyword & Clicks & Metric Day & Ad Campaign\\
    \midrule
     chopper axe & 6 & B00EOVRX06 & 1 & 2023-09-01 & 12345\\
    \bottomrule
  \end{tabular}
\end{table*}

\section{Product-Cluster Randomized Control Trial Design}

To address the challenge of interference in experimental designs, we propose a preemptive modeling strategy that incorporates interference networks during the design phase. This approach allows us to shift the unit of randomization from individual products to clusters of products, as illustrated in Figure 3. Importantly, traditional constrained randomization methods \cite{moulton2004covariate}, such as segmenting by product categories, prove ineffective. This is because search engines can associate broad upper funnel search queries (e.g., "Harry Potter") with a diverse range of products across multiple categories (e.g., a book, toy, or blanket related to Harry Potter). By leveraging interference networks, our method ensures more robust and accurate experimental outcomes.

\setlength{\parskip}{1em}
\textbf{Modeling Network Interference} The notion of interference in the network \cite{hudgens2008toward} we construct has to be aligned with the notion of interference we are trying to estimate. Since the relevance of products to a user search query is determined by the search engine and ranking algorithm, an advertiser cannot use its internal datasets that provide product to keyword mapping e.g. e-commerce website's own search to product results. Instead, we use daily reports provided by the ad publisher itself. These reports have information on which actual user search query on the search engine was mapped to which shopping ads product by the ad publisher. A sample mock row from such a report is shown in Table 1. We use these search query reports to construct an undirected bipartite search query - product graph using the number of impressions as edge weight.

\textbf{Unipartite Projection} To apply one mode projection of the bipartite graph onto the product nodes in order to model the between-network interference, we needed a scoring function to attribute weights to the resulting graph edges. Since, we start with large number of products (~200M+), we could not directly use the edge weighting functions proposed by \citet{stram2017weighted} due to the computational complexity. Instead we propose the following edge weight function that requires signiﬁcantly less computation:

\begin{multline}
W_{\text{uni}}(a, b) = \sum_{i=1}^{n} \frac{1}{\log_e(f_{sq_i})} \\
\frac{\min(W_{\text{bi}}(sq_i, a), W_{\text{bi}}(sq_i, b))}{\max(W_{\text{bi}}(sq_i, a), W_{\text{bi}}(sq_i, b))} \, I[sq_i, a, b]
\end{multline}

where:
\begin{enumerate}
    \item $W_{\text{uni}}(a, b)$ is the edge weight in the unipartite graph (one-mode projection) between product a and b.
    \item $W_{\text{bi}}(sq_i, a)$ is edge weight between search query $i$ and $a$ in the original bi-partite graph.
    \item $f_{sq_i}$ is the number of distinct products that a particular search query drives impressions to. Since the distribution is right-skewed i.e. few upper funnel queries drive impressions to only a few distinct products, weighing down by the log of frequency of search query helped us to weigh down edge weight contributions between two products from very generic queries.
    \item $I[sq_i, a, b]$ is 1 if search query $sq_i$ trigger an impression for both product $a$ and $b$ as represented by the presence of an edge in the original bipartite graph, otherwise 0.
\end{enumerate}

Using the above approach to take a weighted one-mode projection of the bipartite graph leads to an increase in the number of edges, since if a search query links to $n$ products, we need to consider $\binom{n}{2}$ pairs of edges.

\begin{figure}[h]
  \centering
  \includegraphics[scale=0.36]{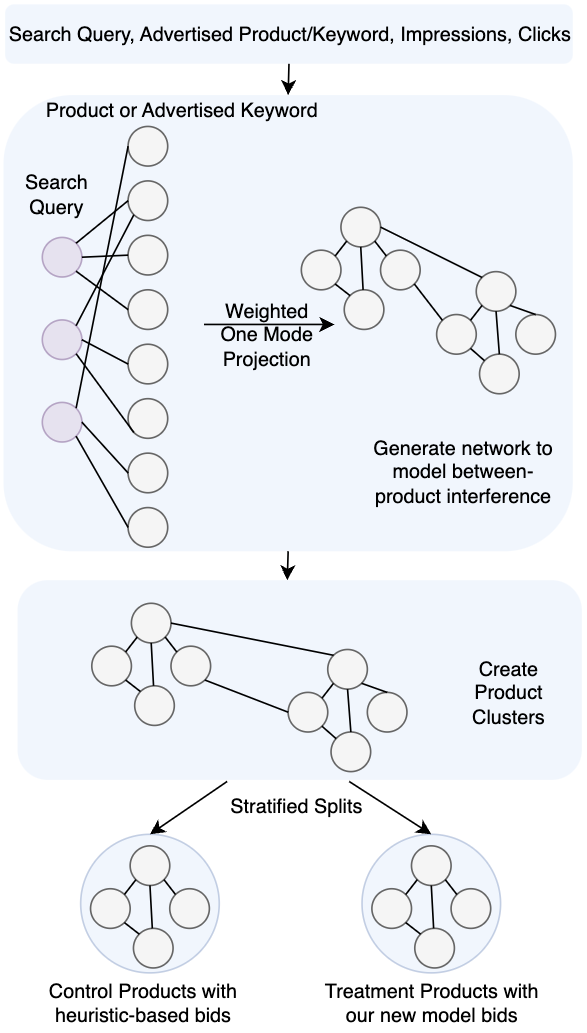}
  \caption{Birds-eye view of our experiment design. First, we fetch the search query reports from the ad publisher and create a search-query, product bipartite graph. Then we take a weighted one mode projection and ﬁnally use a clustering algorithm to cluster products. We then do a stratiﬁed random split of product clusters.}
\end{figure}

\textbf{Graph Partitioning Methodology} Constructing a product graph following the above approach then allows us to use network dismantling algorithms \cite{braunstein2016network} as oppose to naive connected components approach to creating product clusters. Historically, the community identiﬁcation problem is a well studied problem in computer science literature \cite{newman2004detecting, girvan2002community, newman2004fast}. However, a lot of the proposed methods wouldn't scale up since we are dealing with graphs that are as large as 200M+ nodes and 400M+ edges. Thus, anything that runs in $O(|V|^2)$ or $O(|E|^2)$ is not practical. Moreover during the graph partitioning phase, we needed to ﬁnd a balance between two objectives:

\begin{enumerate}
\item Maximize the number of product clusters as they translate to randomization units. More clusters equal more power for our test.
\item Minimize the between clusters edge weights.
\end{enumerate}

Since, the more clusters we create, the less isolated they are, these two objectives are conflicting. For example, if we want to have zero connections across clusters, then the obvious solution is to have one cluster only. This, of course, would not lend itself to an A/B test. To balance the above two objectives, first we look at the percentage edge weight across $k$ clusters ($C_1, ..., C_k$) i.e. leakage as $L$:

\[
L = \frac{\sum_{i=1}^k \sum_{j \in C_i, k \notin C_i} W_{\text{uni}}(j, k)}{\sum_{j, k} W_{\text{uni}}(j, k)}
\]

Secondly, we use a clustering algorithm that is designed for balanced clustering, that is, all clusters should have roughly equal size. We evaluated naive connected components, power iteration clustering (PIC) \cite{lin2010power}, and METIS \cite{karypis1997metis}. Connected components minimizes leakage, but suffers from extreme imbalance. PIC improves cluster imbalance, but suffers from high leakage. We choose to use METIS partitioning algorithm which is an extremely efficient and fast implementation of graph partitioning algorithm for undirected weighted graph. METIS adopts an objective function to minimize the number of weighted edges whose vertices belong to different partitions. The METIS graph partitioning consists of three phases: (i) In the graph coarsening phase, a series of successively smaller graphs is derived from the input graph. This process continues until the size of the graph has been reduced to just a few hundred vertices, (ii) In the initial partitioning phase, a partitioning of the coarsest and hence, smallest, graph is computed and finally (iii) in the un-coarsening phase, the partitioning of the smallest graph is projected to the successively larger graphs by assigning the pairs of vertices that were collapsed together to the same partition. After each projection step, the partitioning is refined using heuristics to iteratively move vertices between partitions as long as such moves improve the quality of the partitioning. The advantages of this methodology are threefold:

\begin{enumerate}
    \item It runs in $O(|E|)$ time, which is extremely efficient for large graphs.
    \item It is the only algorithm that allows precise control of both the number partitions and the balances of the overall split.
    \item It is the only algorithm that is specifically trying to minimize the edgecut (defined as weighted sum of edges that straddle between different clusters).
\end{enumerate}

\textbf{Optimal number of clusters:} We want to be able to identify as many nearly independent clusters as possible with leakage controlled within the tolerance. We plot leakage against various choice of $k$ (number of partitions), and identify a $k$ that is as large as possible where the leakage is as small as possible (i.e. identifying the elbow point). For the new shopping ad bidder experiment, we ended up having 10,000 clusters and 36\% edge weight across clusters. Note, the above measure ($L$) overstates the spillover effects as they consider spillover between clusters that may end up being in the same group (C or T).

\textbf{Magnitude of Spillover:} The search query - product bipartite graph we construct usually has a clustering coefficient [20] of around $\sim0.6$ for most marketplaces which indicates tightly knit groups in the network suggesting high spillover. However, to empirically provide a lower bound on the magnitude of bias due to interference we need to conduct a meta-experiment that randomizes over two experiment designs: one Bernoulli randomized, one cluster randomized. We can then check for a statistically significant difference between the total average treatment effect estimates obtained with the two designs \cite{saveski2017detecting}. In the absence of business approval to run such a meta-experiment, the next best directional data point we have is from our previous attempt to run simple product-split A/B test. The impact measured from that experiment had been largely overstated ($\sim44\%$ lift) when compared to the actual lift ($\sim24\%$ lift) observed.

\section{Application}

In this section, we discuss the use-case motivated in Section 2 to show an application of the product-cluster randomized control trial design. We had developed a new machine learning based product valuation model for our shopping ads program to improve over the current in production heuristic bidding algorithm and we wanted to run an online experiment to understand the impact on the long term profit. We used the methodology described in Section 3 to create product clusters based on search query reports from the ad publisher for the past one year.

\textbf{Constrained Randomization} Once we had the clusters, we created strata of clusters with similar characteristics instead of randomizing them in a simple bernoulli fashion. We measure the net impressions, clicks, cost and profit of each of the product cluster and stratify clusters on those axis.

\textbf{Experiment Setup} The goal of this experiment was verifying the null hypothesis that the new bidding strategy is better than the current bidding strategy in term of bidding efficiency i.e. increase of net long term profitability while maintaining the total ad spend. We matched the spend between control and treatment groups to control for elasticity as well as to comply with spend constraints at account level. Finally, we run a simulation based power analysis for cluster randomized designs using difference-in-differences (DID) estimation \cite{angrist2009mostly}. 

\textbf{Measurement} To measure the impact of the proposed valuation method, a DID analysis for cluster randomized designs is performed for two weeks of periods where spends are closely matched. The results from the DID analysis showed a lift in click-through-rate for the treatment group which was consistent with the lift observed post roll out of the new bidding model. Note that since model errors can be correlated within cluster, failure to control for within-cluster error correlation can lead to misleading small standard error and consequently low p-values. Although we do not control for within-cluster error correlation in the model, post-estimation we obtain cluster-robust standard errors as proposed by \citet{white2014asymptotic}.

\section{System Architecture}

Our product-cluster randomized control methodology as detailed in Section 3 asks for a highly scalable and ﬂexible infrastructure with very different compute requirements and library support for each step. To address these challenges we propose the "Search Marketing Lab" using AWS SageMaker \cite{joshi2020amazon} pipelines which allows to deﬁne a series of interconnected processing steps where each step (i) can be provided its own docker image that has our code in preferred language and (ii) can have its own compute environment. This allows for polyglot programming. Here, we brieﬂy focus on the split generation component. In particular, we break the approach into 3 modules:

\begin{figure}[h]
  \centering
  \includegraphics[width=\linewidth]{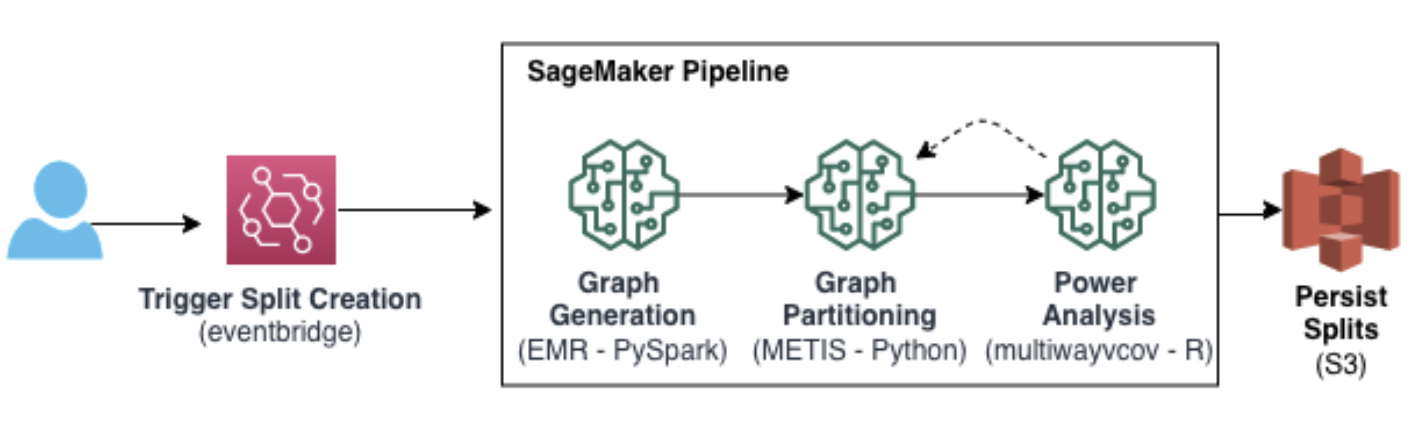}
  \caption{Birds-eye view of Product-cluster split generation system.}
\end{figure}

\begin{enumerate}
    \item \textbf{Graph Generation} which requires parsing >1 year of daily search query report data and Keyword data to build a graph edge list - thus requiring spark’s distributed compute. We use a cluster of memory-optimized instances for this step.
    \item \textbf{Graph Partitioning} In this step we use the METIS graph partitioning algorithm. METIS is written entirely in ANSI C (no distributed implementation) but there is a python wrapper for the METIS library \cite{karypis1997metis} that we use. We use a single compute-optimized instance for this step.
    \item \textbf{Power Analysis} In this module we obtain cluster-robust standard errors after ﬁtting a linear mixed effect model, using R’s cluster.vcov \cite{arellano1987computing} implementation to return a multi-way cluster-robust variance-covariance matrix and perform inference for estimated coefﬁcients using R’s coeftest.
\end{enumerate}

SageMaker Pipeline executions can be scheduled using Amazon EventBridge passing run-time parameters. This allows to deﬁne a single pipeline with multiple executions (e.g. one per marketplace) based on input parameters. The serves as a blueprint for a large scale production system combining multiple languages (R, Python on Spark) utilizing each to their respective strengths (R for statistical analysis modules, python on Spark for ETL) triggering SageMaker processing jobs orchestrated via SageMaker Pipelines.

\section{Conclusion and Future Work}

In this paper, we present a cluster-based randomized control test design which enables search marketing e-commerce teams to do fast online experiment launch while minimizing interference between experimental groups. Our key idea is to use observational data to construct bipartite (Search Query - Product) SERP interference networks and use a novel weight function to take weighted projections to form unipartite graphs which can be use to create clusters of products appearing together on SERP (via Paid Search shopping ads, text ads or Free Search listings), and then using those clusters to randomize on. Online A/B testing results for the treatment group are consistent with the lift observed post roll out of a new bidding model thereby showing that the A/B test design gives a good estimate of the actual lift. In our previous attempts to run simple product-split A/B test the impact measured from experiments had been largely overstated because of spillover effects. Lastly, we present a novel simplified system architecture using SageMaker which allows scientist to do polyglot programming using compute and language suitable for each scientiﬁc module. 

One downside of inferring interference network from search query report data is that such observational data is censored, that is, we only have data when we win the auction. In future, we are investigating using SERP page data from platforms like seoClarity to get better visibility into SERP interference networks and allow us to incorporate not just shopping ad products but also Text Ads keyword and Free Search URLs to build comprehensive ad units spanning across all Search channels - Text Ads, Shopping Ads and Free Search. More recently, this also includes large language models powered results like shown in Appendix. We can than use these ad units to design cross-channel substitution experiments. We are also working on investigating further into the stability of these clusters over time and that they can be updated in real time as more data ﬂow in from search engines. Finally, we are exploring recent proposed experiment designs by \citet{bajari2021multiple} to measure the actual magnitude of spillovers.

\begin{acknowledgments}
   We extend our heartfelt gratitude to Doug Wong and Mike James for their invaluable support and funding, which made this research possible. We also wish to thank Kingshuk RoyChoudhury and Han Wu for their insightful discussions and constructive feedback. Their expertise and thoughtful engagement have greatly contributed to the development and refinement of our ideas.
\end{acknowledgments}

%%
%% Define the bibliography file to be used
\bibliography{sample-ceur}

\appendix

\section{Components of Interest on Recent LLM Powered SERPs}

\begin{figure}[h]
  \centering
  \includegraphics[scale=0.45]{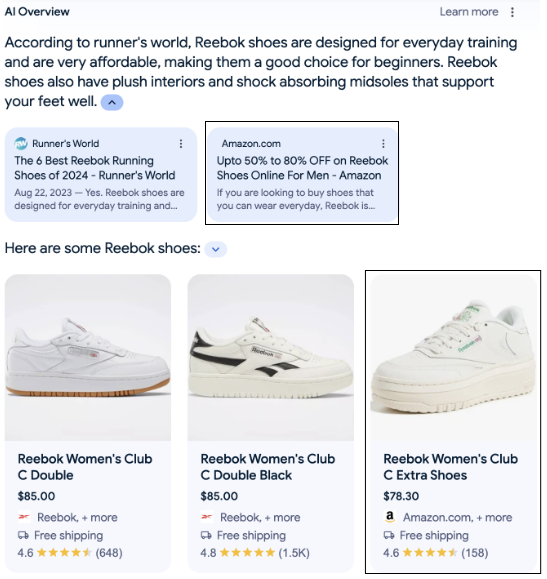}
  \caption{Evolving Large Language Model powered SERPs with components of interest highlighted in rectangle boxes.}
\end{figure}

\end{document}